\documentclass{egpubl}
\usepackage{hpg25DL}

\WsPaper           

\usepackage[T1]{fontenc}
\usepackage{dfadobe}  
\usepackage{amsmath}
\usepackage{subcaption}
\RequirePackage{amssymb}
\usepackage{cite}  
\BibtexOrBiblatex
\electronicVersion
\PrintedOrElectronic
\ifpdf \usepackage[pdftex]{graphicx} \pdfcompresslevel=9
\else \usepackage[dvips]{graphicx} \fi

\usepackage{egweblnk} 

\title[Interactive Stroke-based Neural SDF Sculpting]%
      {Interactive Stroke-based Neural SDF Sculpting}

\author[F. Rubab \& Y. Tong]
{\parbox{\textwidth}{\centering F. Rubab\orcid{0009-0001-6979-7746}
        and Y. Tong\orcid{0000-0002-7929-4333} 
        }
        \\
{\parbox{\textwidth}{\centering Michigan State University \& USA
       }
}
}

\begin{document}

\teaser{
 \includegraphics[width=\linewidth]{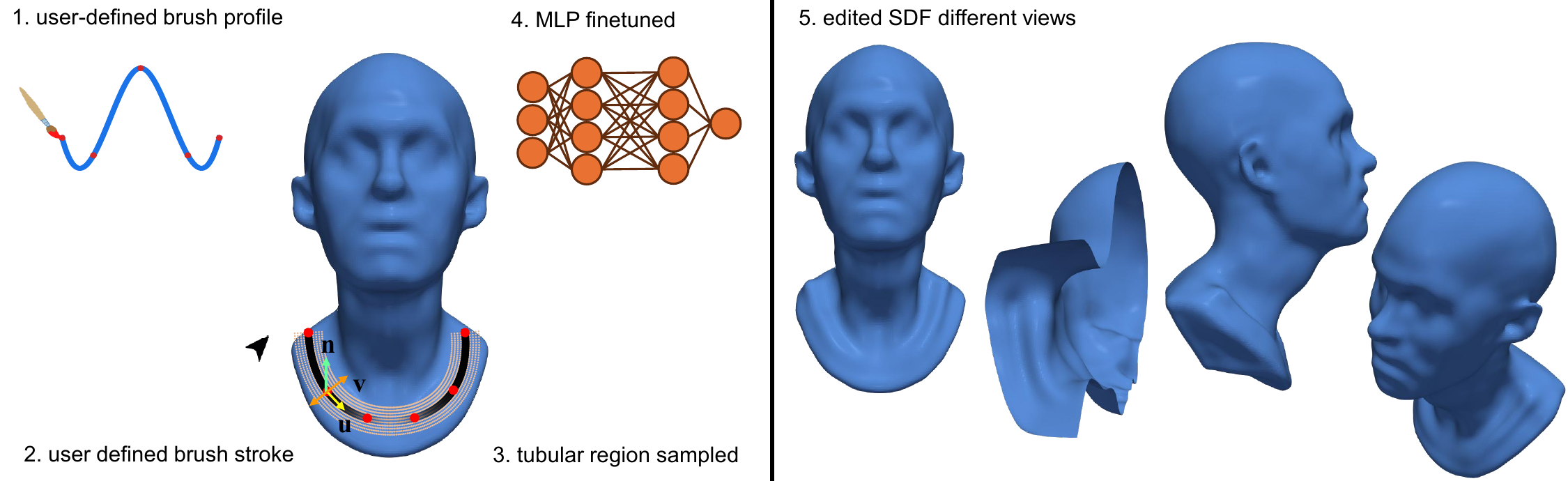}
 \centering
  \caption{Overview of our interactive neural sculpting pipeline. \textit{1}: User defines a custom brush profile by selecting control points in a $[-1,1]$ window, interpolated with a cubic spline. \textit{2}: A brush stroke is drawn on the neural SDF-rendered shape in an interactive viewer with darker regions indicating stronger deformation intensity due to modulation. \textit{3}: A moving $u$-$v$-$n$ coordinate frame and tubular sampling region are established around the stroke path. \textit{4}: Samples are used to query the SDF’s MLP for fine-tuning. \textit{5}: Resulting deformations, in this case, realistic collarbones, are shown from multiple angles (front, cutaway, side and isometric views).}
\label{fig:overview}
}

\maketitle
\begin{abstract}
Recent advances in implicit neural representations have made them a popular choice for modeling 3D geometry. However, directly editing these representations presents challenges due to the complex relationship between model weights and surface geometry, as well as the slow optimization required to update neural fields. Among various editing tools, sculpting stands out as a valuable operation for the graphics and modeling community. While traditional mesh-based tools like ZBrush enable intuitive edits, a comparable high-performance toolkit for sculpting neural SDFs is currently lacking. We introduce a framework that enables interactive surface sculpting \emph{directly} on neural implicit representations with optimized performance. Unlike previous methods, which are limited to spot edits, our approach allows users to perform stroke-based modifications on the fly, ensuring intuitive shape manipulation without switching representations. By employing tubular neighborhoods to sample strokes and customizable brush profiles, we achieve smooth deformations along user-defined curves, providing intuitive control over the sculpting process. Our method demonstrates that versatile edits can be achieved while preserving the smooth nature of implicit representations, all without compromising interactive performance.

\ccsdesc[300]{Computing methodologies~Computer graphics}
\ccsdesc[300]{Computing methodologies~Machine learning}

\printccsdesc
\end{abstract}  
\section{Introduction}
\begin{figure*}
    \centering
    \includegraphics[width=0.9\linewidth]{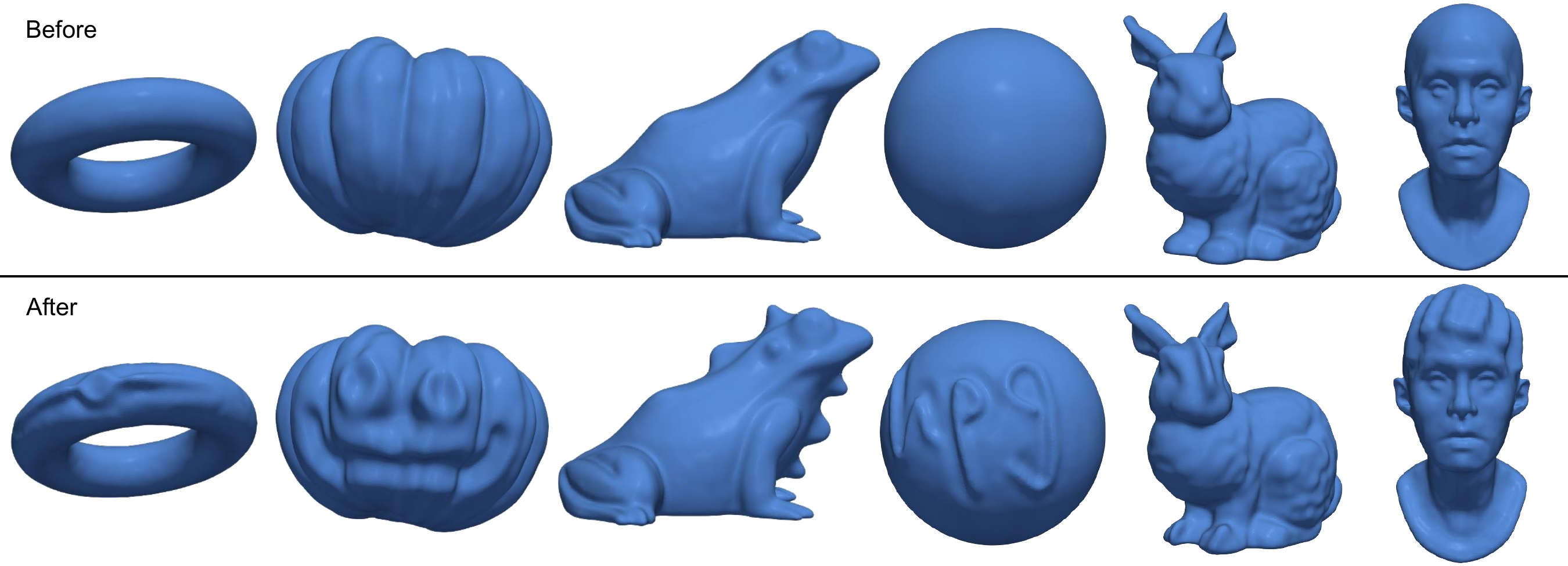}
    \caption{Examples of natural artistic edits using our method. Each column shows a sample shape from the dataset, with the original SDF on the first row and the edited version below. (a) Ring with a central gem and tapering carvings. (b) Halloween-themed pumpkin carving. (c) Frog morphed into a lizard with scales using a sine-modulated stroke. (d) 'hpg' inscription on a sphere with an asymmetric brush. (e) Bunny transformed with robotic features on its face, feet, and ears. (f) Victorian hairstyle added to a previously bald bust.}
    \label{fig:edits}
    \vspace{-0.4cm}
\end{figure*}
Shape modeling and manipulation are essential across fields like computer graphics, animation, Computer-Aided Design (CAD), and virtual reality, fueling decades of research. Traditionally, shapes have been represented using explicit methods like Bézier curves, splines, and polygonal meshes, which allow direct manipulation of control points or vertices for flexible 3D modeling.

Over time, shape representations in practice expanded to also include point clouds, voxel grids, and volumetric models. Among these, Signed Distance Functions (SDFs) \cite{bloomenthal1997introduction, osher2004level} became popular for Constructive Solid Geometry (CSG) \cite{foley199612}, offering a compact way to define complex geometries by level set functions that measure distances to surfaces.
Recently, implicit neural representations (INRs) have gained attention for encoding 3D shapes continuously and compactly using Multilayer Perceptrons (MLPs) \cite{sitzmann2020implicit}. Neural Signed Distance Functions (neural SDFs), in particular, allow a network to estimate the SDF of a shape, enabling sampling at arbitrary resolution and offering efficient storage for complex geometries \cite{park2019deepsdf, atzmon2020sal, atzmon2020sald}. However, neural implicit representations are difficult to edit; unlike meshes or splines, where vertices or control points allow direct adjustments, neural SDF parameters are embedded within network weights globally, making local modifications challenging. This lack of interactive editability in neural SDFs limits their applicability.

While some efforts attempt to enhance neural SDF editability by modifying shape codes or latent spaces \cite{deng2021deformed}, they often lack fine control and immediate feedback. In contrast, traditional sculpting tools allow intuitive, real-time surface manipulation, essential in fields like animation and design. However, such an interactive toolkit is yet to be realized for neural implicit representations.

In this work, we introduce an expressive framework that addresses the challenge of editing 3D shapes represented by neural SDFs. Our method provides a comprehensive, interactive toolkit for sculpting and carving shapes, offering users responsive control over surface modifications. Unlike prior approaches that focus on limited point edits with significant latency \cite{tzathas20233d}, our work supports stroke-based edits, allowing users to define arbitrary brush profiles and apply them smoothly across surfaces in a responsive manner.

Our approach leverages a performance-conscious tubular sampling strategy to create smooth, continuous deformations along a user-defined stroke. By focusing samples within the interaction region, each neural update involves fewer points, making optimization significantly faster and enabling responsive feedback and intuitive control. In addition, the flexibility of the framework supports complex, user-defined brush profiles and modulation along the stroke path, enabling a wide range of specialized edits and enhancing the creative potential of artists and designers. With this toolkit, complex shapes can be both represented and edited within the same framework, eliminating the need to switch between different representations. \footnote{\tiny Code: \url{https://github.com/Fizza-Rubab/Interactive-Neural-Sculpting}}

To summarize, our main contributions include:
\begin{itemize}
\item We introduce an interactive toolkit for editing neural SDFs with stroke-based capabilities tailored for 3D modeling.
\item Our framework allows for custom brush profiles with modulation along the stroke path, supporting effects like chiseled strokes, asymmetric shapes, and simultaneous carving and extrusion.
\item We present efficient tubular sampling for stroke-based deformations in neural SDFs, achieving up to 15× speedup over prior point-based methods, resulting in interactive user feedback during editing.
\end{itemize}

\section{Related Work}
\subsection{Implicit Neural Representations}
Also known as neural fields, INRs have recently emerged as a powerful approach for representing continuous signals that can be queried at arbitrary spatial resolution \cite{Xie2021NeuralFields}. Parameterized by MLPs, INRs map input coordinates directly to field values, making them resolution-independent and providing an alternative to traditional grid-based representations that rely on discrete sampling. Before neural networks, implicit functions were widely used in graphics and vision, such as CSG with analytic SDFs \cite{iquilezles} or fractal generation using complex equations \cite{mandelbrot1983fractal}. Neural networks expanded their applicability, as demonstrated in Neural Radiance Fields (NeRF) \cite{mildenhall2021nerf}, which model radiance and density fields to synthesize realistic views, and DeepSDF \cite{park2019deepsdf}, which represents 3D geometry via continuous SDFs parameterized by MLPs.

Despite their effectiveness, INRs struggle with high-frequency details due to the low-frequency bias of MLPs, leading to oversmoothing in complex textures and fine-grained geometry. Several methods address this limitation. Positional encoding, originally from transformers \cite{vaswani2017attention}, and Fourier features \cite{tancik2020fourier} lift input coordinates into a higher-dimensional space to improve high-frequency representation. SIREN instead employs sinusoidal activation functions to encode high-frequency information without requiring input transformation \cite{sitzmann2020implicit}.

INRs are also computationally intensive, particularly in real-time applications. Research has explored techniques such as neural hash maps and adaptive sampling to optimize efficiency while preserving fidelity \cite{muller2022instant}. KiloNeRF \cite{reiser2021kilonerf} speeds up NeRF \cite{mildenhall2021nerf} by partitioning the scene into a grid of small MLPs, each responsible for a localized region, significantly reducing training and inference time.

Neural fields that parameterize geometry are central to our work. DeepSDF \cite{park2019deepsdf} represents a shape's SDF with an MLP and generalizes to shape classes using latent codes. Occupancy Networks follow a similar approach but learn a continuous occupancy probability field \cite{mescheder2019occupancy}. SAL \cite{atzmon2020sal} discards the sign and instead learns an unsigned distance function, while SALD \cite{atzmon2020sald} refines this by incorporating derivative information in the loss. Other methods integrate the eikonal property of SDFs as a loss term \cite{yang2024stabilizing}. Beyond loss-based improvements, some works introduce alternative supervision strategies, such as using 2D images instead of 3D point clouds. For instance, \cite{bangaru2022differentiable} employs reparameterization techniques for differentiable end-to-end optimization from image supervision, achieving high-quality geometry reconstruction.

\subsection{Shape Editing Techniques}
Shape editing has been a key area in geometric modeling, evolving from explicit surface-based methods to deep-learning approaches. Traditional techniques relied on explicit geometry representations like meshes and NURBS (Non-Uniform Rational B-Splines) \cite{versprille1975computer}, allowing precise vertex and control point manipulation. Methods such as Free-Form Deformation \cite{sederberg1986free} and Skeletal Animation/rigging enabled localized transformations without manual vertex adjustments. More advanced techniques, like vector field-guided deformations \cite{von2006vector}, leverage vector fields to control shape transformations, while multi-resolution editing \cite{zorin1997interactive} allows modifications at different levels of detail.

\subsection{Interactive Sculpting}
Our work addresses a specialized form of editing in the modeling community: digital sculpting, which enables artists to shape, carve, push, and pull a 3D model to create intricate forms. Typically, sculpting is achieved through user-defined strokes using virtual brushes that vary in intensity, size, and shape, allowing for highly detailed and expressive edits. For explicit representations, there are numerous tools, the most notable of which include Blender \cite{blender} and ZBrush \cite{zbrush}, offering an intuitive and highly interactive sculpting experience.

For implicit representations, however, interactive sculpting options are more limited. Some CSG-based applications, such as the open-source tool MagicaCSG \cite{magicacsg} and the web-based platform Womp \cite{womp}, provide basic shape editing through Boolean operations. 

For neural implicit representations, the only relevant work we found for sculpting on a surface represented as a zero level-set is 3D Neural Sculpting (3DNS) \cite{tzathas20233d}, which serves as our baseline by enabling interactive editing of neural SDF. 3DNS provides a framework for interactively editing neural SDFs via point-based modifications on a surface's zero-level isoset. Users sculpt the surface using analytic polynomial radial brushes, each defined by radius and intensity. These brushes are smooth, positive 2D functions defined over a unit disk, reaching a maximum at the origin and tapering to zero at the unit circle. To apply an edit, 3DNS selects sample points on a disk tangent to the interaction point, then projects them onto the unaltered surface, and finally shifts them perpendicularly by a distance defined by the brush function. The system modifies the surface only within the disk area, creating localized edits based on the brush template and radial distance from the interaction point. For surface sampling, 3DNS balances model-preserving samples outside the disk and interaction samples within the disk using a weighting scheme. For model-preserving samples, a Markov chain process is used to ensure more uniform sample distribution on the surface. 

While 3DNS achieves point-based edits, it lacks flexibility in brush profiles and the more typical stroke-based deformations in sculpting. Its operations approximate heat kernel deformations and are confined to additive unions of point edits, making continuous, modulated strokes impractical. Attempting to replicate a stroke with sequential point edits is computationally expensive, as each point incurs separate local sampling and surface projection, resulting in redundant evaluations and large optimization batches. This not only hinders interactivity but also makes precise control difficult—and in some cases impossible—due to overlapping brush influences and uneven sampling density.

\section{Method}
\begin{figure}
  \centering
    \includegraphics[width=\linewidth]{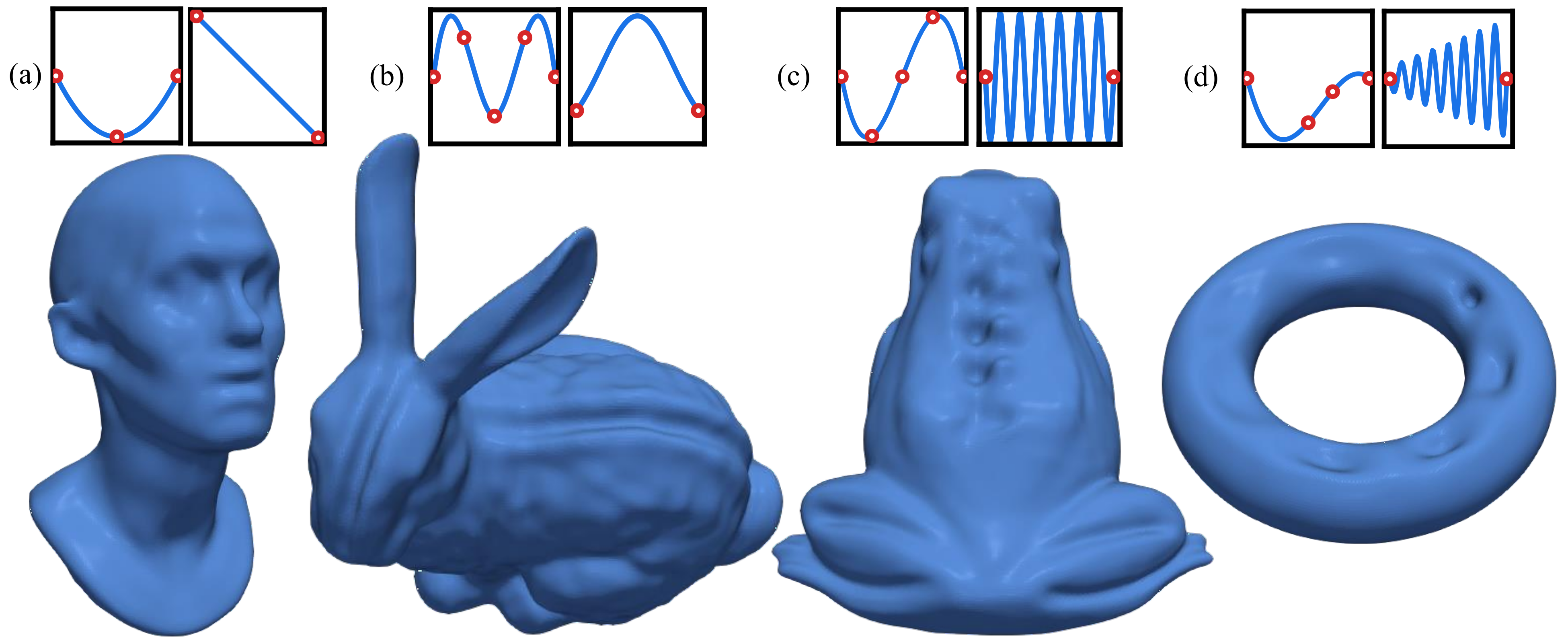}
    \caption{Examples of edits using different brush profiles and modulation functions. Above each shape, the left plot shows the brush profile, and the right plot shows the modulation.}
    \label{fig:vary}
  \vspace{-5pt}
\end{figure}

We present an interactive neural sculpting framework built around pretrained neural SDFs, allowing real-time stroke-based edits on 3D surfaces. The system operates by sampling points along a user-defined stroke, forming a tubular region around the curve where edits are applied. Central to this approach is the brush profile, which defines the shape and intensity of the deformation. Users can either select from analytical brush functions, such as linear, cubic, or quintic falloffs, or define custom profiles using control points, which are interpolated with splines.

Additionally, the brush profile can be modulated along the stroke, enabling effects like gradual offset distance changes, fade in, fade out, etc. The combination of tubular sampling and efficient surface evaluation allows for smooth and consistent deformations, while maintaining interactive update speed through lightweight fine-tuning of the neural SDF during editing.

\subsection{Neural SDF}
The backbone system of our approach is an implicit neural representation that models 3D geometry using a signed distance function. Neural SDFs represent surfaces implicitly by parameterizing the geometry using a neural network, and effectively allow continuous sampling of 3D shapes at arbitrary resolution without being tied to any fixed grid.

\subsubsection{Signed Distance Function}
Given a surface \( S \subset \mathbb{R}^3 \), the signed distance function assigns to every point \( x \in \mathbb{R}^3 \) the shortest distance to the surface. Specifically, for a closed surface, it is defined as
\begin{equation}
\text{SDF}(x, S) =
\begin{cases}
   \min_{y \in S} \; d(x, y) & \text{if } x \text{ is inside } S, \\
  - \min_{y \in S} \; d(x, y) & \text{otherwise},
\end{cases}
\end{equation}
where \( d(x, y) \) is the Euclidean distance between points \( x \) and \( y \).

A neural SDF uses a neural network to approximate this function, effectively capturing the surface as the zero-level set of the network's output.

\subsubsection{SIREN based MLP}

For the neural SDF representation, we adopt the SIREN architecture \cite{sitzmann2020implicit}, a simple MLP with sinusoidal activations that enable the modeling of high-frequency details. SIREN takes spatial coordinates as input and outputs the corresponding signed distance value, with sine activations helping to represent fine surface structures more effectively than standard activations like ReLU.  

The non-linear layers use sine functions:  
\begin{equation}  
\text{SIREN}(x) = \sin(\omega_0 \cdot W x + b),  
\end{equation}  
where \( W \) is the weight matrix, \( b \) is the bias, and \( \omega_0 \) controls the frequency of the sine waves.

While alternative architectures could improve sharpness and better preserve unedited regions, our focus is not on optimizing the neural SDF backbone but on developing an efficient stroke-based sculpting pipeline on top of the INR. Thus, we adopt SIREN, following 3DNS, to ensure direct comparability while maintaining simplicity and effectiveness.

\subsubsection{Neural SDF Editor}
During interactive editing, the current state of the implicit surface is rendered using a raymarching pipeline. A regular grid of rays is cast from the camera, and intersections with the zero-level set of the neural SDF are computed via iterative sphere tracing. Normals are estimated from SDF gradients, and shading is applied using a simple Phong illumination model. The resulting image is written to a screen-space texture and displayed. The display is refreshed upon editing or moving the camera around. Meshes for export or offline visualization can optionally be generated using Marching Cubes after editing. Mouse clicks are recorded to define stroke points for sampling the interaction region, while keyboard controls allow the user to adjust brush intensity, radius, modulation mode, and profile.

\subsubsection{Training and Loss Function}
Learning an SDF is a regression problem that necessitates minimizing a loss function that adequately represents a smooth shape. During training, this involves sampling points within the shape, outside the shape, and on the surface within its bounding box, evaluating the loss at each iteration. For surface points, the SDF value is zero, representing the zero-level set. For other sampled points, the loss evaluation requires computing the distance to the nearest surface point, which can be computationally expensive. To address this, we leverage the fact that the SDF satisfies the eikonal equation, adopting a pseudo-loss term based on the eikonal condition for all non-surface points. Additionally, we introduce a term that aligns the normal vectors at these sampled points with the expected surface normals. Finally, a regularization term is included to penalize small SDF values for points near the edges of the bounding box.

This multi-component loss function, adopted from the framework established in 3DNS~\cite{tzathas20233d} can be expressed as
\begin{equation}
L = \lambda_1 L_{\text{regression}} + \lambda_2 L_{\text{eikonal}} + \lambda_3 L_{\text{normal}} + \lambda_4 L_{\text{boundary}},
\end{equation}
where \(L_{\text{regression}}\) represents the primary regression loss, \(L_{\text{eikonal}}\) ensures compliance with the eikonal equation, \(L_{\text{normal}}\) aligns the normals, and \(L_{\text{boundary}}\) enforces regularization near the bounding box edges:
\begin{align}
&L_\text{regression}(\theta) &&= \mathbb{E}_{p_S} \left[|f_\theta(x)|\right]\\
 & L_\text{normal}(\theta) &&= \mathbb{E}_{p_S} \left[g(\nabla_x f_\theta(x), n_x)\right]\\
  &  L_{\text{eikonal}}(\theta) &&= \mathbb{E}_q \left[\left| \|\nabla_x f_\theta(x)\| - 1 \right| \right]\\
   & L_{\text{boundary}}(\theta) &&= \mathbb{E}_q \left[ e^{-\alpha |f_\theta(x)|} \right]
\end{align}
Here \( \lambda_1=1.5\times 10^3, \lambda_2=5, \lambda_3=2.5, \) and \( \lambda_4=5 \) are balancing weights, \( \alpha=100 \) is a large positive constant, $\mathbb{E}_{p_S}$ represents expectation computed over points on surface and $\mathbb{E}_{q}$ for the off-surface points in the bounding box, $\theta$ is the network weights, $f_\theta$ is the SIREN network, $g$ is the cosine distance, and $n_x$ is the surface normal at $x$.  These terms collectively ensure that the network learns an accurate and smooth representation of the surface, while maintaining the correct geometry for off-surface points.

The sampling strategy for training, including the efficient surface sampling used for stroke-based edits, will be discussed in Sec.~\ref{sec:sampling}. 
\subsection{Coordinate Frame for Stroke-Based Sculpting}
To enable intuitive sculpting, we define a custom coordinate system along each stroke that allows for precise control over deformation. This coordinate frame is established with three orthogonal directions:
\begin{enumerate}
    \item \textbf{Stroke Direction (\(u\))}: Defines the main path of the stroke, representing the curve or line along which the brush moves. 
   \item  \textbf{Brush Profile Direction (\(v\))}: Runs perpendicular to \(u\) and tangential to the surface, and defines the extent of the brush profile at any given point along the stroke.
   \item  \textbf{Normal Direction (\(n\))}: Orthogonal to both \(u\) and \(v\), pointing outward from the surface, allowing deformations to be applied in line with the surface geometry.
\end{enumerate}

This \(u\)-\(v\)-\(n\) frame allows the brush profile to be defined along \(v\), while modulation along the stroke itself is achieved along \(u\), and deformations are applied in the normal direction \(n\), as illustrated in Fig.~\ref{fig:overview}.

\begin{figure}[htbp]
  \centering
   \includegraphics[width=0.3\linewidth]{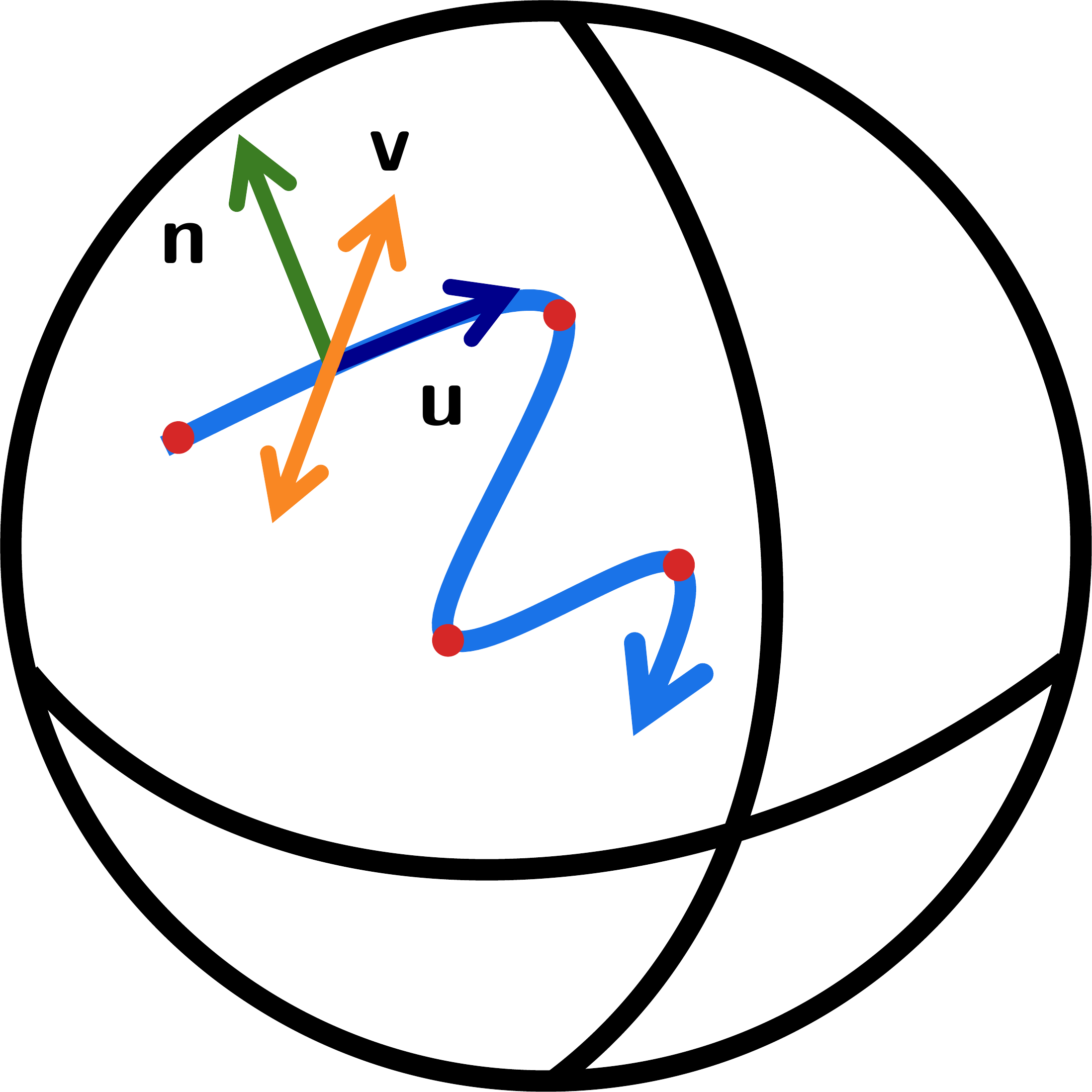}
   \includegraphics[width=0.4\linewidth]{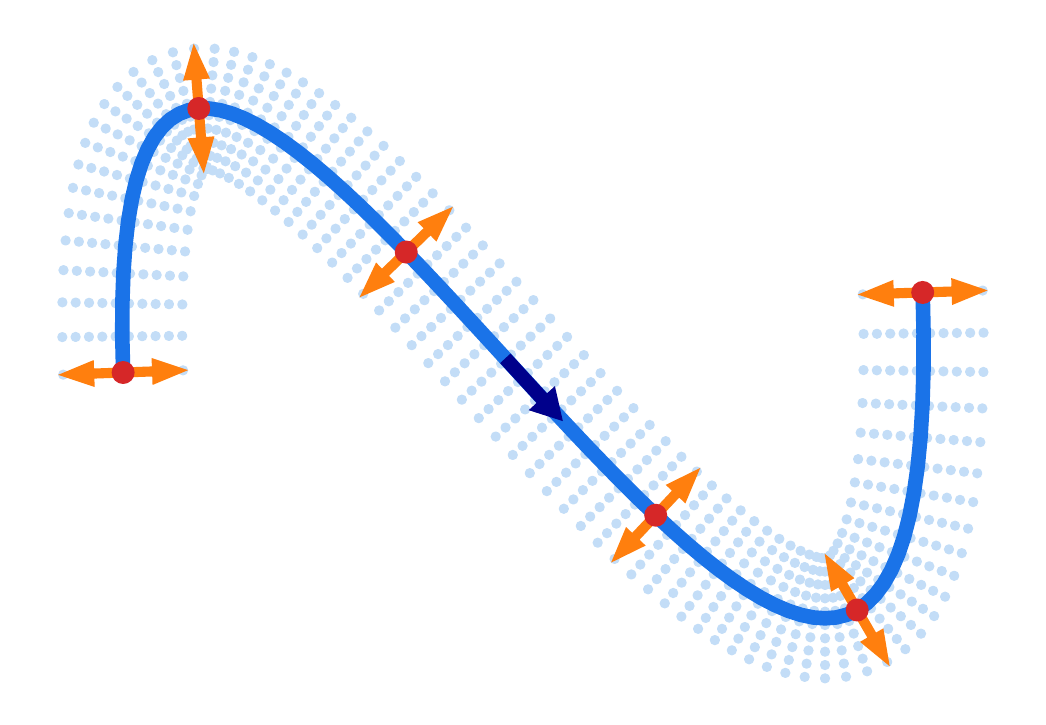}

   \caption{Illustration of the coordinate frame and sampling strategy. \textit{Left}: A 3D view of the user-defined stroke with control points and the $u$-$v$-$n$ frame on the surface. \textit{Right}: A top-down view of the $u$-$v$ plane showing tubular sampling within a radius around the stroke.}
   \label{fig:sampling}
   \vspace{-0.4cm}
\end{figure}

\subsection{Brushes}

Once the neural SDF is trained, the user can perform surface edits using a \textit{brush}. In this framework, a brush is defined as a \(C^1\) function, \(b_P(v)\), operating along the \(v\)-direction. The brush profile determines the intensity (amount of normal displacement) and shape of the deformation across the stroke, and it can be defined by users interactively using control points. We normalize the domain and range of the brush to be within $[-1, 1]$. Such a brush definition is flexible, allowing for both carving and extruding operations within a single application as the displacement can be negative as well as positive. Specifically, we use the piecewise polynomial function \(P(v)\), defined by the Catmull-Rom spline \cite{catmull1974class} interpolating the $k$ user-defined control points \((v_i, y_i)\) for \(i = 1, 2, \ldots, k\),
\begin{equation}
b_P(v)=P(v), \quad |v| \le 1,
\end{equation}
as illustrated in Fig.~\ref{fig:overview}.

Given the normalized brush profile \(b_P(v)\), following 3DNS~\cite{tzathas20233d}, we define a family of brushes \(B_{r,s}(x)\) parameterized by radius \(r\) and intensity \(s\):

\begin{equation}
B_{r,s}(x) = s \; b_P\left( \frac{x}{r} \right), \quad r \in \mathbb{R}^+, \, s \in \mathbb{R}.
\end{equation}

\begin{figure}[htbp]
  \centering
    \includegraphics[width=\linewidth]{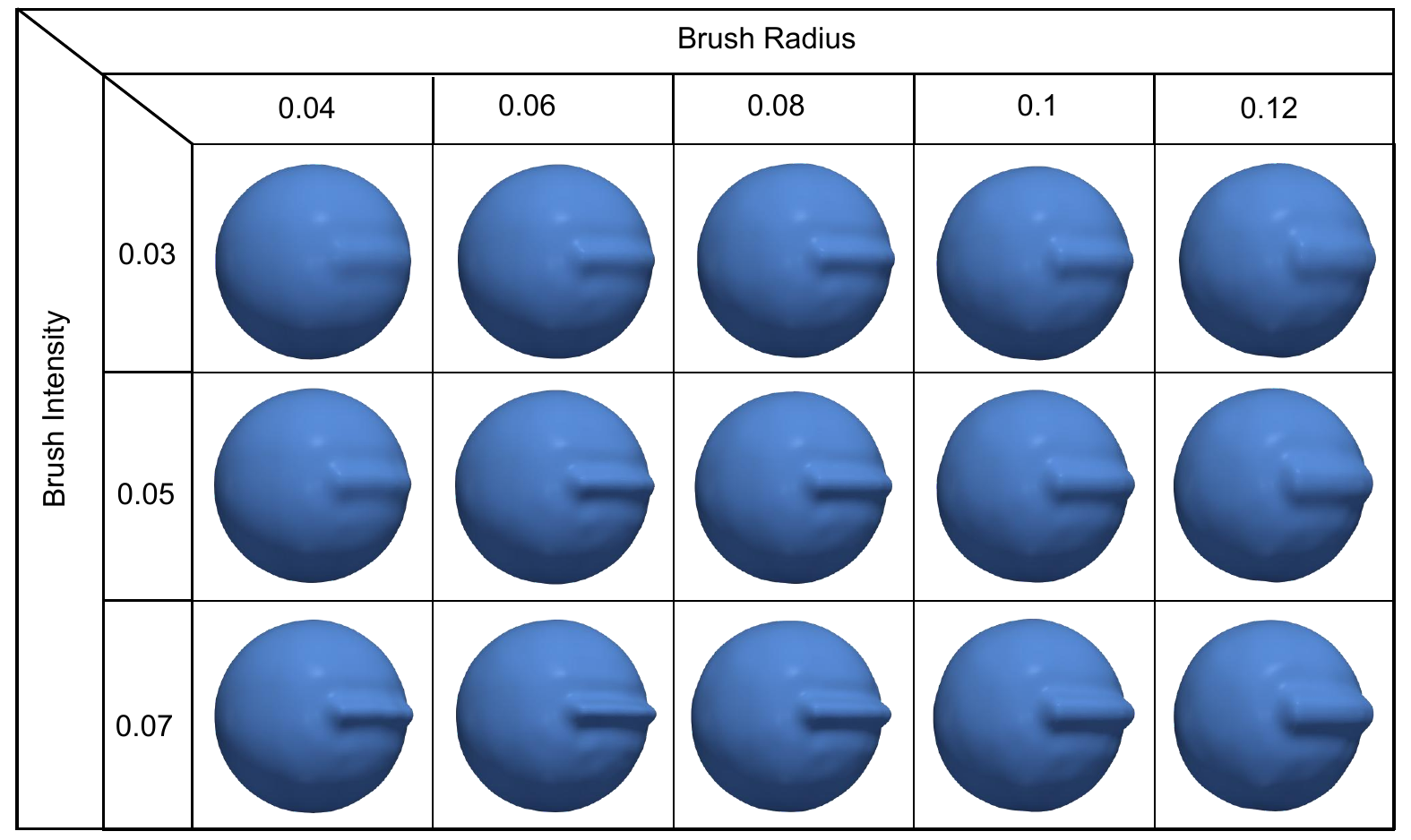}
    \caption{Effect of radius and intensity variations for the same brush stroke.}
    \label{fig:brushfamily}
\end{figure}

Since the brush profile has normalized domain and range, this formulation enables independent adjustments of the region and magnitude of deformation. For instance, given a positive user-defined brush profile, a positive intensity \(s\) can create a bump on the surface, whereas a negative intensity can result in a dent with a radius controlling the width of the impact region of the edit. Fig. \ref{fig:brushfamily} shows the effect of different intensities and radii when applying a parabolic brush stroke on a sphere. 

To apply the brush, we modify the neural SDF’s zero-level set \( f_{\theta}(p) = 0 \) by offsetting surface points along the normal direction using the brush profile within the local \(u\)-\(v\)-\(n\) frame. Points \( p \) are first sampled in the tubular region around the stroke and projected onto the surface to ensure they lie on the original zero-level set. Their displaced positions are then computed as:  
\begin{equation}  
p' = p + B_{r, s}(v(p)) \; n(u(p)),  
\end{equation}  
where \( p' \) is the target zero-level set position, \( v(p) \) is the signed distance from \( p \) to the brush stroke curve, measured in the plane orthogonal to \( u \), with the sign indicating the relative position to the plane formed by the surface normal and stroke direction. \( n(u(p)) \) is the surface normal at the stroke. The neural SDF is then fine-tuned to align its zero-level set with the deformed surface by minimizing:  
\begin{equation}  
L_\text{deformation} = \mathbb{E}\left[ \left| f_{\theta}(p') \right| \right],  
\end{equation}  
where \(f_{\theta}(p')\) is the modified SDF at \(p'\) and \( \mathbb{E} \) represents the mean absolute SDF value over all displaced points \( p' \), ensuring they remain on the zero-level set of the updated surface.

\begin{figure*}
  \centering
   \includegraphics[width=0.9\linewidth]{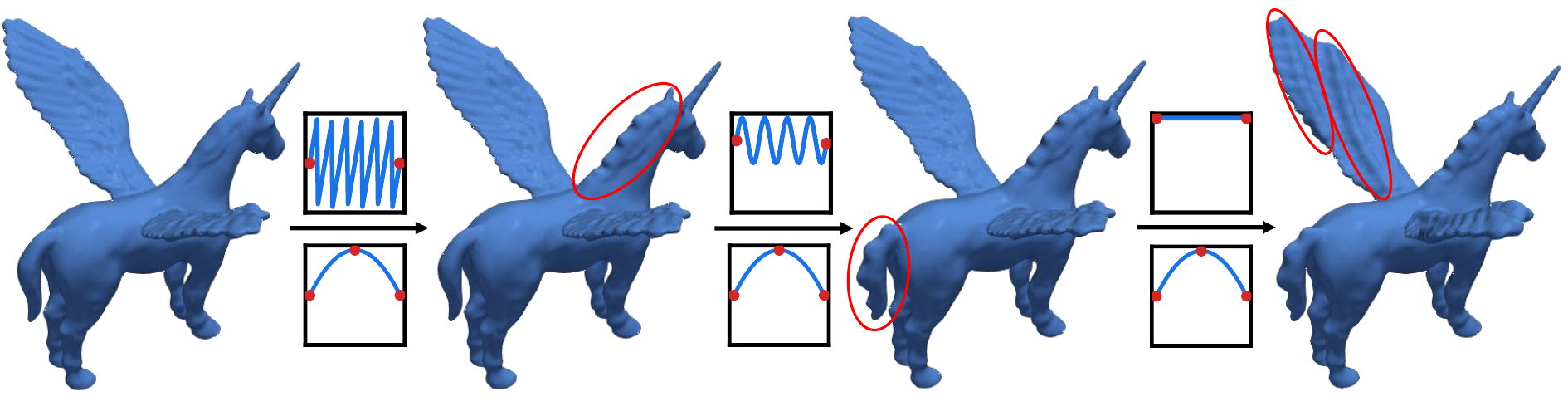}

   \caption{Editing sequence on a flying-unicorn model. For each step, the plot above the arrow shows the modulation curve, while the plot below shows the brush profile. The user successively sculpts tufts on the mane, stylizes the tail, and adds decorative ridges to the wings.}
   \label{fig:unicorn}
\end{figure*}
\subsection{Stroke Representation}

In our framework, a stroke is defined as a curve with parameter $u$, representing the trajectory of the brush center moving on the 3D surface. The tangent at a point on the curve provides the $u$-direction. This path, or stroke, can be directly specified by the user by selecting a sequence of control points on the surface through the interactive editor. These control points are interpolated using a cubic spline to create a smooth, continuous curve $\gamma(u)$, parameterized by $u \in [0,1]$, where $u=0$ represents the start and $u=1$ the end of the stroke.

Mimicking typical sculpting edits, the intensity of deformation can vary dynamically along the stroke’s $u$-direction through a modulation function $m(u)$. Modulation strategies can be user-defined or selected from predefined functions, allowing for varied effects along the stroke. For instance, a linear falloff reduces intensity gradually from one end to the other, a central modulation peaks in the middle of the stroke and tapers toward both ends, while a sinusoidal modulation creates a rhythmic intensity pattern such as frog scales in Fig. \ref{fig:edits}. Custom user-defined modulation curves are also supported, offering artists the flexibility to create personalized intensity variations along each stroke path. 

At each point along the stroke, this modulation function \(m(u)\) combines with the brush profile \(B_{r, s}(v)\) in the perpendicular \(v\)-direction to yield a spatially dynamic deformation field. The overall brush intensity at any point is therefore expressed as 
\begin{equation}
m(u) \; B_{r, s}(v),
\end{equation}
where \(B_{r, s}(v)\) is the brush profile perpendicular to the stroke. Since the modulation operates directly on the stroke parameterization, it introduces expressive control without incurring additional computational cost. Fig. \ref{fig:edits} presents naturalistic edits that artists would find useful. Fig. \ref{fig:vary} highlights editing capabilities of our method using different brush profiles and modulation functions: (a) a linearly fading cheekbone, (b) central modulation of a three-way brush on the bunny’s spine and head, (c) a non-radial sinewave-modulated braid on the frog’s back, and (d) dampened sinewave detailing on a torus. 

\begin{table*}[htbp]
  \centering
  \begin{minipage}{0.23\textwidth}
  
    \centering
    \setlength{\tabcolsep}{2.5pt}
    \begin{tabular}{|cccc|} 
    \hline
    \multicolumn{4}{|c|}{Editing time (sec)}                                                                               \\ \hline
    \multicolumn{1}{|c|}{Points} & \multicolumn{1}{c|}{Ours}    & \multicolumn{1}{c|}{3DNS}     & Gain              \\ \hline
    \multicolumn{1}{|c|}{8}             & \multicolumn{1}{c|}{0.468} & \multicolumn{1}{c|}{0.553}  & \textbf{1.183} \\ \hline
    \multicolumn{1}{|c|}{16}            & \multicolumn{1}{c|}{0.473} & \multicolumn{1}{c|}{1.083}  & \textbf{2.292}  \\ \hline
    \multicolumn{1}{|c|}{32}            & \multicolumn{1}{c|}{0.464} & \multicolumn{1}{c|}{1.985}  & \textbf{4.28}  \\ \hline
    \multicolumn{1}{|c|}{64}            & \multicolumn{1}{c|}{0.482} & \multicolumn{1}{c|}{3.767}  & \textbf{7.808}  \\ \hline
    \multicolumn{1}{|c|}{128}           & \multicolumn{1}{c|}{0.517} & \multicolumn{1}{c|}{7.755} & \textbf{15.011} \\ \hline
    \end{tabular}
    \caption{Comparison of editing times for our method versus 3DNS method, averaged over 100 iterations. Gain represents the relative speedup achieved by our method.}
      \label{tab:timing}
    \vspace{-1em}
  \end{minipage}
    \hfill
  \begin{minipage}{0.56\textwidth}
    \centering
    \setlength{\tabcolsep}{2.5pt}
    \begin{tabular}{|ccccccc|}
    \hline
    \multicolumn{7}{|c|}{Mean   Chamfer Distance $\times 10^3 (\downarrow)$}                                                                                                                                               \\ \hline
    \multicolumn{1}{|c|}{}        & \multicolumn{3}{c|}{Over whole surface}                                                             & \multicolumn{3}{c|}{Inside tubular region}                                       \\ \hline
    \multicolumn{1}{|c|}{Shape}   & \multicolumn{1}{c|}{Ours}           & \multicolumn{1}{c|}{3DNS}  & \multicolumn{1}{c|}{Coarse Mesh} & \multicolumn{1}{c|}{Ours}            & \multicolumn{1}{c|}{3DNS}   & Coarse Mesh \\ \hline
    \multicolumn{1}{|c|}{Sphere}  & \multicolumn{1}{c|}{\textbf{7.279}} & \multicolumn{1}{c|}{7.394} & \multicolumn{1}{c|}{7.329}       & \multicolumn{1}{c|}{\textbf{9.932}}  & \multicolumn{1}{c|}{28.099} & 17.532      \\ \hline
    \multicolumn{1}{|c|}{Bunny}   & \multicolumn{1}{c|}{\textbf{9.175}} & \multicolumn{1}{c|}{9.190} & \multicolumn{1}{c|}{9.815}       & \multicolumn{1}{c|}{\textbf{14.661}} & \multicolumn{1}{c|}{29.130} & 21.240      \\ \hline
    \multicolumn{1}{|c|}{Bust}    & \multicolumn{1}{c|}{\textbf{7.363}} & \multicolumn{1}{c|}{7.555} & \multicolumn{1}{c|}{7.617}       & \multicolumn{1}{c|}{\textbf{12.711}} & \multicolumn{1}{c|}{29.164} & 19.998      \\ \hline
    \multicolumn{1}{|c|}{Pumpkin} & \multicolumn{1}{c|}{\textbf{8.470}} & \multicolumn{1}{c|}{8.690} & \multicolumn{1}{c|}{9.445}       & \multicolumn{1}{c|}{\textbf{9.672}}  & \multicolumn{1}{c|}{28.446} & 20.301      \\ \hline
    \multicolumn{1}{|c|}{Torus}   & \multicolumn{1}{c|}{\textbf{5.969}} & \multicolumn{1}{c|}{6.098} & \multicolumn{1}{c|}{6.208}       & \multicolumn{1}{c|}{\textbf{11.850}} & \multicolumn{1}{c|}{23.368} & 15.307      \\ \hline
    \multicolumn{1}{|c|}{Frog}    & \multicolumn{1}{c|}{\textbf{8.099}} & \multicolumn{1}{c|}{8.357} & \multicolumn{1}{c|}{9.091}       & \multicolumn{1}{c|}{\textbf{8.785}}  & \multicolumn{1}{c|}{28.262} & 22.886      \\ \hline
    \end{tabular}
    \caption{Comparison of our editing method with 3DNS pointwise edits and direct mesh editing on a coarse mesh of equivalent network size. Chamfer distances were computed using 100,000 points, with means averaged over 10 independent edits per shape.}
    \label{tab:chamfer}
    \vspace{-1em}
  \end{minipage}%
  \hfill
  \begin{minipage}{0.14\textwidth}
    \centering
    \setlength{\tabcolsep}{4pt}
    \begin{tabular}{|c|c|}
                \hline
                $n$ & CD ($\downarrow$) \\
                \hline
                8   & 0.01252 \\\hline
                16  & 0.01149 \\\hline
                32  & 0.00967 \\\hline
                64  & 0.00674 \\\hline
                128 & 0.00654 \\\hline
                512 & 0.00642 \\\hline
            \end{tabular}
            \caption{Effect of varying number of point samples along the stroke on edit quality.}
            \label{tab:cd_values}
    \vspace{-1em}
  \end{minipage}
\end{table*}
\subsection{Sampling}
\label{sec:sampling}

Sampling is essential for efficiently capturing both stroke-affected and untouched regions of the surface for training the neural parameters $\theta$. The main sampling coordinates $(u,v,n)$ define the structure of the brush interaction.

\subsubsection{Tubular Sampling}

A key contribution of this paper is the \textit{tubular sampling strategy} we employ for efficient stroke-based edits. To apply the brush profile along the curve, samples are generated not only along the stroke itself (the \( u \)-direction) but also around it in a tubular neighborhood defined by the perpendicular \( v \)-direction. This strategy ensures consistent brush effects across the region with an appropriate brush size, even on intricate, curved surfaces.
 
To perform tubular sampling, we first generate \(N \) samples uniformly distributed along the stroke trajectory \( \gamma(u) \). For each point on the curve, we calculate an in-plane normal vector \( \mathbf{v}(u) \) by taking the cross product of the tangent vector \( \mathbf{u}(u) \) along the curve with the surface normal \( \mathbf{n}(u) \):
\begin{equation}
\mathbf{v}(u) = \mathbf{n}(u) \times \mathbf{u}(u).
\end{equation}

Next, we apply uniformly spaced offsets in \( v \)-direction from \([-r, r]\), where \( r \) is the brush radius. These offset points are then projected onto the surface to capture the tubular neighborhood, creating a continuous zone of interaction for the brush. Fig.~\ref{fig:sampling} demonstrates our sampling strategy applied to a stroke.

Based on our experiments, we find that sampling \textbf{99 points} along the stroke, \textbf{101 points} in the offset direction and 10,000 samples to stabilize the untouched region effectively captures a typical stroke on the surface, yielding smooth deformations at interactive frame rates. It is significantly more efficient than 3DNS, which samples 120,000 surface points and discards those in the interaction region. It then resamples 10 (referred to as the interaction factor) times the number of discarded points within the tangent disc around each stroke point. When using 10,000 surface samples within the interaction region for both methods, 3DNS typically samples more than 10 times as many overall points as our method. Our compact sampling directly translates to smaller optimization batches, improving update speed while preserving edit fidelity.

\section{Experiments}

In this section, we evaluate our proposed neural sculpting framework with continuous stroke-based edits and custom brush profiles. We test the performance of our method on multiple 3D objects and compare it with the point-edit approach used in previous work, as well as with traditional mesh-based editing. We use Chamfer distance as the primary metric to compare the geometric differences between the original and edited models. Additionally, we compare the editing speed of our approach with 3DNS. 

\subsection{Implementation Details}
All experiments were conducted on an NVIDIA GeForce RTX 4070 GPU (8GB) using a codebase implemented in Python with PyTorch \cite{paszke2019pytorch}. For geometry operations, \texttt{trimesh} \cite{trimesh2022} and \texttt{Open3D} \cite{zhou2018open3d} are used. The interactive interface, SDF rendering, and visualization are implemented with \texttt{PyOpenGL} \cite{pyopengl}, a Python wrapper for OpenGL. Profiling was performed using PyTorch’s built-in profiler. 

\subsection{Dataset and Testing Setup}
For our experiments, we use the same dataset as \cite{tzathas20233d}. This dataset comprises six 3D shapes: a frog, bust, and pumpkin (sourced from TurboSquid \cite{turbo}), the Stanford Bunny \cite{turk1994zippered}, and two analytical models—a sphere with radius 0.6 and a torus with major radius 0.45 and minor radius 0.25. These serve as the base models for our edits, e.g., in Fig~\ref{fig:edits}. The mesh models were preprocessed by normalizing the coordinates to ensure they fit within a bounding box $[-1,1]$, with additional space reserved for editing. We follow the same preprocessing steps as in \cite{tzathas20233d} to make the comparison consistent. We represent these shapes using a neural SDF parameterized by a compact MLP, specifically using the SIREN architecture with 2 hidden layers and 128 neurons per layer.

\subsection{Editing Performance Metric}
To quantitatively assess the accuracy of our edits, we use the chamfer distance, a widely used metric in geometric processing tasks. Given two point clouds, \(A\) and \(B\), their Chamfer distance is defined as:

\begin{equation}
d_{\text{chamfer}}(A, B) = \frac{1}{|A|}\sum_{a \in A} \min_{b \in B} \|a - b\| + \frac{1}{|B|}\sum_{b \in B} \min_{a \in A} \|b - a\|
\end{equation}

This measures how close the edited surface is to the ground truth by summing the nearest point distances between the two point clouds. Chamfer distance is computed over the entire surface as well as within the interaction region where edits are applied.

\subsection{Editing Comparison}

In this section, we compare our method with three baselines: edits on a high-resolution ground truth mesh, a low-resolution mesh, and 3DNS's point-based editing extended to strokes.

For the ground truth comparison, we apply the brush profile to mesh vertices within a radius of curve segments, providing an ideal edit by directly deforming the mesh geometry. We also compare our method to a simplified mesh, which has approximately the same number of triangles as the neural SDF model’s parameters via quadratic decimation, ensuring a fair comparison in terms of representation complexity. Lastly, we extend the point-based editing approach from 3DNS, applying the brush at multiple points along the stroke, represented by a Catmull-Rom spline.

We compute the Chamfer distance over 100,000 samples between the edited surface and the ground truth mesh. For these experiments, the brush radius is 0.08 and its intensity is 0.06. For a fair comparison, a quintic brush profile is used, which can be expressed in the 3DNS framework. 12 point samples are used along the stroke and the number of samples for regularization of the untouched region is set to 120,000. Models are fine-tuned for 100 epochs. The mean is computed over 10 independent random edits and summarized in Table.~\ref{tab:chamfer}, considering both the interaction region and the entire surface. Our method achieves lower error in both cases for all shapes in the dataset.

\begin{figure}[h]
    \centering
    \begin{subfigure}{0.24\columnwidth} 
        \includegraphics[width=\linewidth]{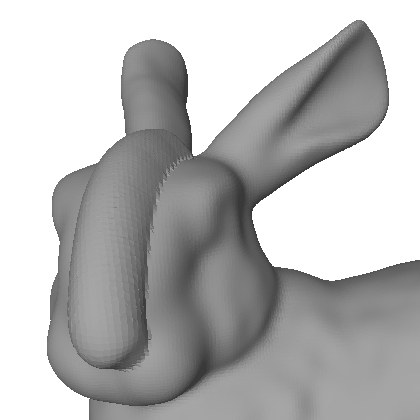}
    \end{subfigure}
    \vline
    \begin{subfigure}{0.24\columnwidth}
        \includegraphics[width=\linewidth]{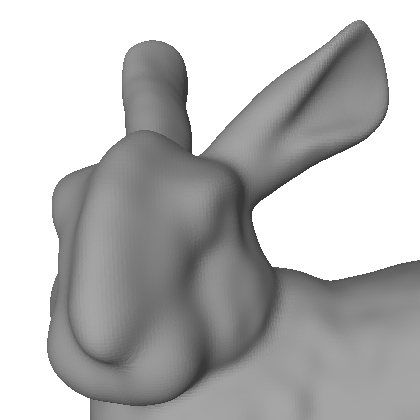}
    \end{subfigure}
    \begin{subfigure}{0.24\columnwidth}
        \includegraphics[width=\linewidth]{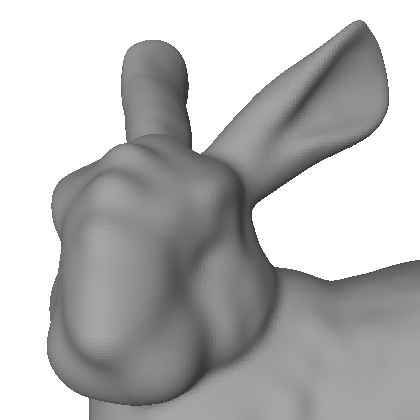}
    \end{subfigure}
    \begin{subfigure}{0.24\columnwidth}
        \includegraphics[width=\linewidth]{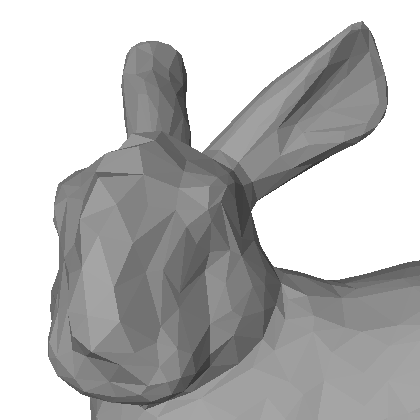}
    \end{subfigure}
    
    \caption{Example of a stroke applied using different methods. From left to right: ideal edit (ground truth), our method, 3DNS, and edit on a coarse mesh.}
    \label{fig:comparison}
    \vspace{-0.4cm}
\end{figure}

Fig.~\ref{fig:comparison} compares the application of a brush stroke across different methods at a similar computational cost. Our approach smoothly approximates the target edit even with sparse sampling along the stroke direction, whereas 3DNS results in jagged deformations due to pointwise editing. The coarse mesh edit, on the other hand, is both noisy and inaccurate.  

Increasing the number of points along the stroke and offset directions improves edit fidelity, as shown by decreasing chamfer distance values in Table~\ref{tab:cd_values} for a fixed sphere edit. While sharp boundary features observed at the edges of the brush profile in Fig.~\ref{fig:comparison} ground truth edit are theoretically achievable with our method, their realization depends on whether the underlying neural representation can support such high-frequency details.

\begin{figure*}
  \centering
   \includegraphics[width=\linewidth]{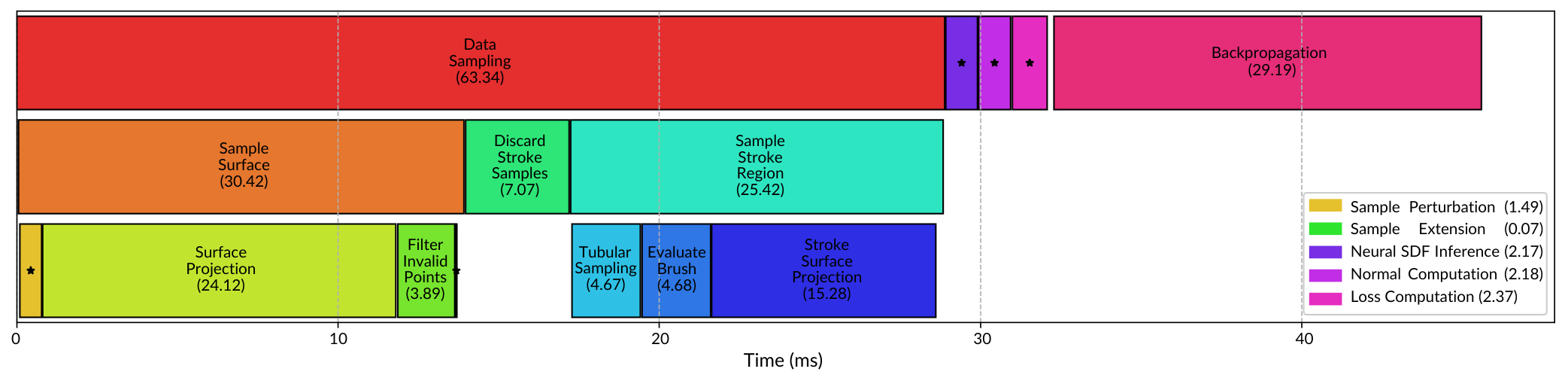}

   \caption{Execution timeline for a single fine-tuning iteration. Each bar represents a recorded stage in the training pipeline, positioned and scaled according to its start time and duration.}
   \label{fig:gantt}
   \vspace{-0.5cm}
\end{figure*}
\subsection{Efficiency Comparison}
The primary performance metric for an editing framework is its speed and interactivity. Table \ref{tab:timing} compares the average time and speedup achieved with our tubular sampling method versus the 3DNS approach, for varying numbers of points sampled along the stroke. The models were fine-tuned for 50 epochs--sufficient to yield a decent edit, with 10,000 sample points used to regularize the model's untouched regions. Results are averaged over 100 iterations. Our tubular sampling method enables faster edits (under a second) by using fewer sample points in the interaction region. In contrast, the point-based approach slows significantly as stroke point density increases, making it impractical for interactive sculpting workflows.

\subsection{Performance Profiling}
To assess the computational cost of our method, we profiled the training loop using PyTorch’s built-in profiler. Table~\ref{tab:profiling-structured} reports the average duration of each stage, measured across 50 training epochs (excluding the first and last to avoid initialization and finalization bias). As expected, the majority of time is spent in the Data Sampling phase, which includes both surface sampling from the current model and interaction-region sampling around brush strokes. The cost is roughly evenly split between these two sources. Within this phase, surface projection—used to snap sampled points onto the zero level set—dominates due to repeated forward and backward passes through the network. Model evaluation and optimization are comparatively efficient, with backpropagation being the most time-consuming component in that segment. Figure~\ref{fig:gantt} shows a Gantt-style execution timeline from a representative median epoch.

\begin{table}[htbp]
\centering
\begin{tabular}{|l|r|}
\hline
\textbf{Stage} & \textbf{Time (ms)} \\
\hline
\multicolumn{2}{|l|}{{Data Sampling}} \\
\hline
\quad Sample Surface              & 13.009 \\\hline
\quad\quad Sample Perturbation   & 0.959 \\
\quad\quad Surface Projection    & 8.980 \\
\quad\quad Post-Projection Filtering & 2.491 \\
\quad\quad Sample Extension      & 0.047 \\\hline
\quad Discard Stroke Samples      & 2.514 \\\hline
\quad Sample Stroke Region        & 13.133 \\\hline
\quad\quad Tubular Sampling      & 1.820 \\
\quad\quad Brush Template Evaluation & 2.089 \\
\quad\quad Stroke Surface Projection & 8.789 \\
\hline
\multicolumn{2}{|l|}{{Model Evaluation}} \\
\hline
\quad Neural SDF Inference        & 1.277 \\\hline
\quad Normal Computation          & 1.223 \\
\hline
\multicolumn{2}{|l|}{{Optimization}} \\
\hline
\quad Loss Computation            & 1.602 \\\hline
\quad Backpropagation             & 11.701 \\
\hline
\end{tabular}
\caption{Breakdown of average durations for each step of our pipeline during a training iteration.}
\label{tab:profiling-structured}
\end{table}

\section{Limitations and Discussion}

\begin{figure}[htbp]
  \centering
   \includegraphics[width=0.9\linewidth]{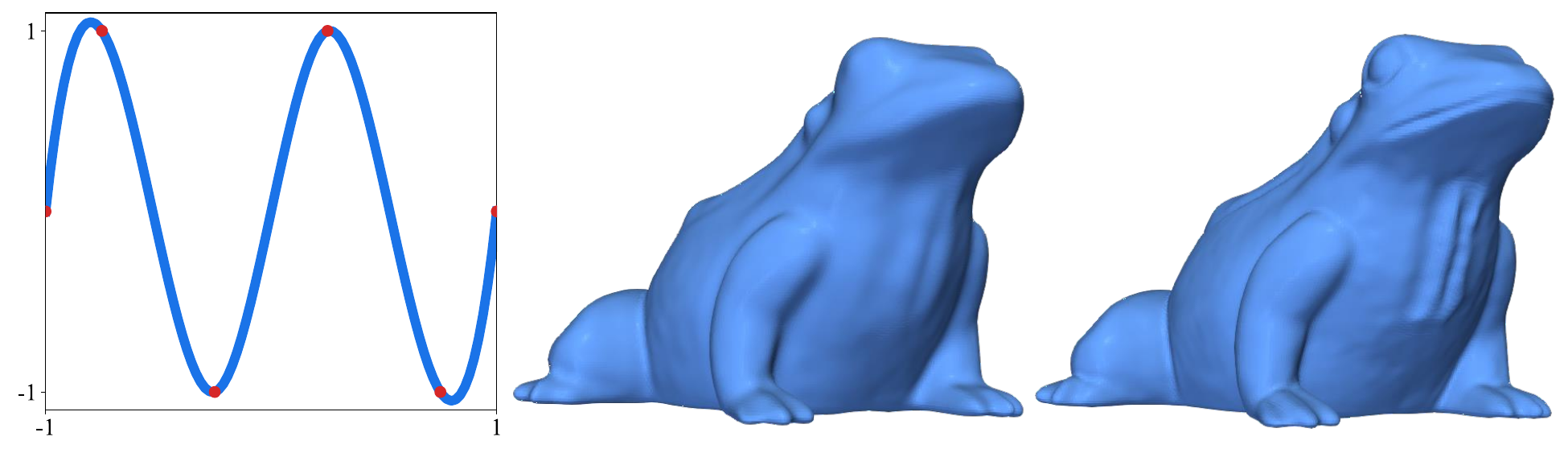}

   \caption{Impact of model resolution on brush template fidelity. Low-resolution model (middle) blurs high-frequency brush oscillations, while the high-resolution model (right) preserves finer details.}
   \label{fig:brush}
   \vspace{-0.5cm}
\end{figure}

Our tool supports neural sculpting on complex shapes as well. To illustrate this, Figure~\ref{fig:unicorn} shows an editing sequence applied to a detailed flying unicorn model from the Objaverse-XL~\cite{objaverseXL} dataset. Edits are applied sequentially to the mane, tail and wings to achieve the desired appearance, demonstrating the system’s applicability to artist-driven workflows. While our method offers versatile sculpting capabilities with fine-grained user control, it has limitations, particularly with highly oscillatory brush profiles. It handles sharp, abrupt changes well (e.g., box-like or linear brushes) but struggles with high-frequency oscillations like wavelets with multiple peaks and troughs, leading to blurring. This trade-off between resolution and complexity in the backbone neural network requires deeper, more expressive MLPs, increasing training and inference costs. Fig.~\ref{fig:brush} shows the same edit on an MLP with more layers. This edit is impossible with 3DNS, even with dense sampling, due to overlapping point influences. Beyond sharpness issues due to their continuous nature, global MLP-based representations also struggle to preserve unedited regions. Limited regularization causes bumpy artifacts, while stronger regularization suppresses edits and increases editing time by requiring more off-target samples. While our sculpting operator is agnostic to the underlying representation, future work could explore architectures with built-in spatial locality, such as wavelet-based MLPs or neural SDF intersections, to improve sharpness and region preservation.

Secondly, very high-intensity brushes can cause abrupt, discontinuous changes in the geometry by inducing large updates to the network weights. This is less of a concern in typical sculpting workflows, where low to medium-intensity edits are generally preferred for achieving smooth, controlled modifications. Thus, while these limitations are worth noting, they do not significantly impede the utility of the framework for most sculpting tasks.

Finally, our implementation is built on PyTorch, which provides flexibility for rapid prototyping but is known to be suboptimal for small networks and batch sizes due to Python overhead and lack of kernel fusion. A custom CUDA implementation with fused kernels could offer significantly higher performance. Additional improvements could be achieved through mixed-precision training and model compilation. For this task in particular, techniques to hide training latency—such as overlapping sampling, performing background updates, or using asynchronous execution could greatly improve responsiveness.

\section{Conclusion}
We present an interactive sculpting framework for efficient stroke-based editing of 3D geometry represented by \emph{neural} signed distance functions. By establishing a tailored coordinate system, formulating brush and stroke representations, and leveraging a performance-conscious tubular sampling strategy, our approach supports smooth, controlled deformations with responsive feedback. This work bridges traditional 3D editing workflows with neural implicit representations, expanding the applicability of neural SDFs in both artistic and scientific domains. Future work may explore region-weight localization and reversible editing to further enhance precision and expressiveness in neural-based 3D design.

\bibliographystyle{eg-alpha-doi} 
\bibliography{egbibsample}       


\end{document}